# Mathematical Model of Age Aggression

Golovinski P. A.

Voronezh State University of Architecture and Civil Engineering

20-letiya Oktyabrya Street, 84, Voronezh, 394006 Russia

e-mail: golovinski@bk.ru

We formulate a mathematical model of competition for resources between representatives of different age groups. A nonlinear kinetic integral-differential equation of the age aggression describes the process of redistribution of resources. It is shown that the equation of the age aggression has a stationary solution, in the absence of age-dependency in the interaction of different age groups. A numerical simulation of the evolution of resources for different initial distributions has done. It is shown the instability of the system and the existence of regimes leading to a concentration of resources in the certain age groups.

Various mathematical models have been formulated and successfully used in application to biology and human society. They demonstrate nontrivial behavior of the systems as a result of variation of parameters [1]. One of widely used models is the known Lottki-Volterra model, describing a competition of two species [2, 3]. The periodic dynamic modes existing in them, is typical for these models [3]. The large successes have been achieved to describe dynamics of complex economical and social systems by computer simulation [4, 5] and with qualitative mathematical analyses of nonlinear systems [3]. Not only some specific scenarios have been discovered, but the limits for parameters of sustainable development were determined too.

The plenty of factors might influence on social and economic dynamics and against one's will we need to develop more coarse schemes, which approximately take into account real correlation in a system. The present work is devoted to the analysis of an age competition in human communities. We started from the point of view that age competition is natural property of the given community. It means, that members of society struggle for limited resources that they have.

In thise relation every person exists in two aspects: on the one hand the person is the source of aggression [6] and he suppresses others, because he tends to redistribute resources to his own favor and, on the other hand, these persons act on suppressively, and restrict his resources. In subsequent discussion, we assume that pair interaction is the prevailing interaction between persons in the process of the age aggression, i.e. this is person-person interaction.

Let us describe the result of age struggle with the help of an age distribution function $W(\tau,t)$, where $\tau$ is the age of society members, $t$ is current time. Among the all-possible modes of the resource distribution dynamics we can try to select steady-state mode with stationary distribution function. Then we get

$$W(\tau,t) = W(\tau) \qquad (1)$$

and dynamics is described by the equation

$$\frac{\partial W(\tau,t)}{\partial t} = 0 \qquad (2)$$

Actually the distribution function can vary in time due to the interaction of persons. We would like to describe intensity of capture and loss of resources by the age functions $h(\tau)$ and $f(\tau)$. To take into account pair interaction we write master-equation in the form

$$\frac{\partial W}{\partial t} = Wh(\tau)\int Wf(\tau_1)d\tau_1 - Wf(\tau)\int Wh(\tau_1)d\tau_1 \qquad (3)$$

Thus receipts of resources in some age group is proportional to the current resources available to the given age group, its ability for resource accumulation, and integrated ability of other groups for losing and their own resources.

Respectively, the outflow of resources is proportional to the current resources available at the given age group, because precisely these resources can be loosed, and it is proportional to the integrated ability to capture resources by other groups too. The Eq. (3) is nonlinear integro-differential master-equation.

It can readily be imagined the qualitative form of $h(\tau)$ and $f(\tau)$ dependencies. The function of the capture activity $h(\tau)$ increases with age up to some maximum value and then decreases to old age. The function of intensity of resource loss $f(\tau)$ decreases at the beginning of life to some minimum value and then it increases. Both functions are essentially positive. The qualitative form of functions $h(\tau)$ and $f(\tau)$ is shown in Fig. 1.

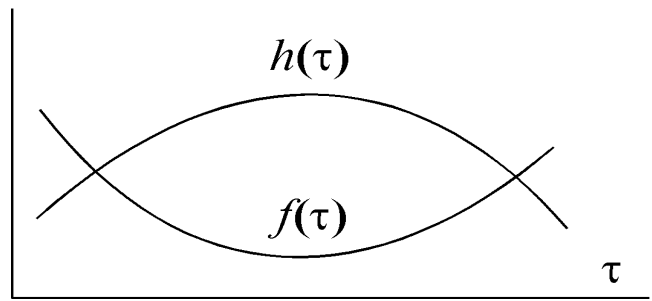

Fig. 1. Activity functions.

After integration of both sides of Eq. (3) with respect to the age parameter we have

$$\frac{\partial \overline{W}}{\partial t} = 0, \qquad (4)$$

and $\overline{W} = \int W \, d\tau = \text{const}$. This fact means, the norm is constant in time and it can be accepted as equal to unit. In this way, the model describes resource redistribution inside community, where the total resource is constant.

The problem of existence of stationary solutions in the model has fundamental importance. After substitution of solution in the form of Eq. (1) in Eq.(3) one can see that left side of equation is equal to zero. Hence

$$\frac{h(\tau)}{f(\tau)} = \frac{\int W(\tau_1,t)h(\tau_1)d\tau_1}{\int W(\tau_1,t)f(\tau_1)d\tau_1} \tag{5}$$

It is evident, that right side of Eq. (5) does not depend on age parameter $\tau$. At the same time the left side of Eq. (5) does not depend on time $t$.

It is possible only when both sides of Eq. (5) are equal to the same constant:

$$\frac{h(\tau)}{f(\tau)} = c \quad \text{and} \quad \frac{\int W(\tau_1,t)h(\tau_1)d\tau_1}{\int W(\tau_1,t)f(\tau_1)d\tau_1} = c \tag{6}$$

Hence $h(\tau)$ is proportional to $f(\tau)$. In particular, for constant total resource the ratio of integrals in Eq. (6) and the value $h(\tau)/f(\tau)$ are equal to unit. In the case of complemented functions we have $h(\tau)+f(\tau)=\text{const}$, and stationary solution exist for $h(\tau)=f(\tau)=\text{const}$ only. It is difficult to obtain some analytical and even approximate solution to the Eq. (3) because of its nonlinearity. At the same time computer simulation for master-equation (3) is rather simple. Starting from initial distribution function $W(\tau,0)$ and some definite functions $h(\tau)$ and $f(\tau)$, we calculate variation of function $W(\tau,t)$ for short time interval $\Delta t$:

$$\begin{aligned} W(\tau,t+\Delta t) &= W(\tau,t) + \frac{\partial W(\tau,t)}{\partial t}\Delta t = \\ &= W(\tau,t) + \left[Wh(\tau)\int Wf(\tau_1)d\tau_1 - Wf(\tau)\int Wh(\tau_1)d\tau_1\right]\cdot\Delta t \end{aligned} \tag{7}$$

Then step-by-step we reproduce the system evolution during finite time interval. In Fig. 2 we show evolution of initial uniform distribution due to the competition processes, as a result of which the basic part of resources concentrated at 60-years age.

Another initial distribution, for example Gaussian function, rather fast tends to the same form that takes place in previous case during the evolution. The dynamics of resource redistribution is most sensitive to the form of the resource loss intensity.

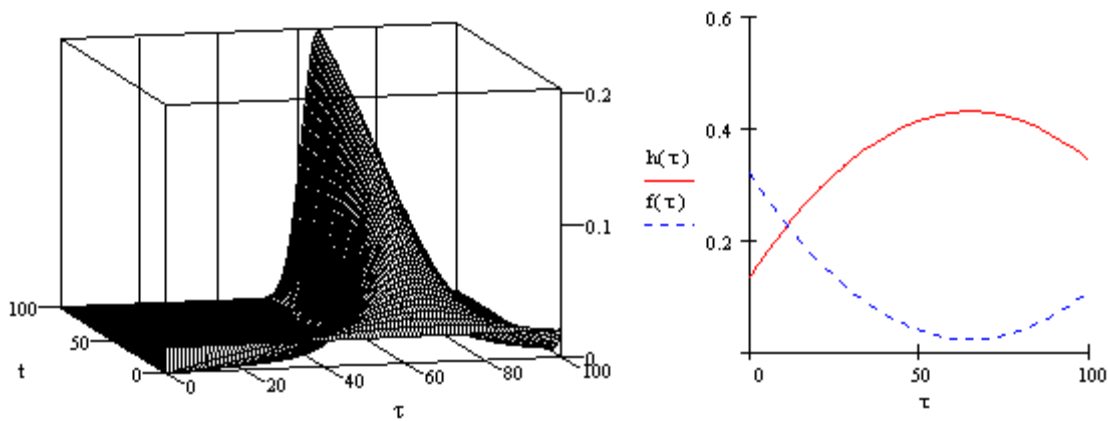

Fig. 2. Redistribution of resources due to the competition processes concentrated at 60-years age.

In Fig. 3 we demonstrate scenario of the resource redistribution that tends to concentration near 40-years age group. Our model of the age aggression obviously demonstrates instability of resource distribution when we take fixed functions of the capture activity and the loss intensity.

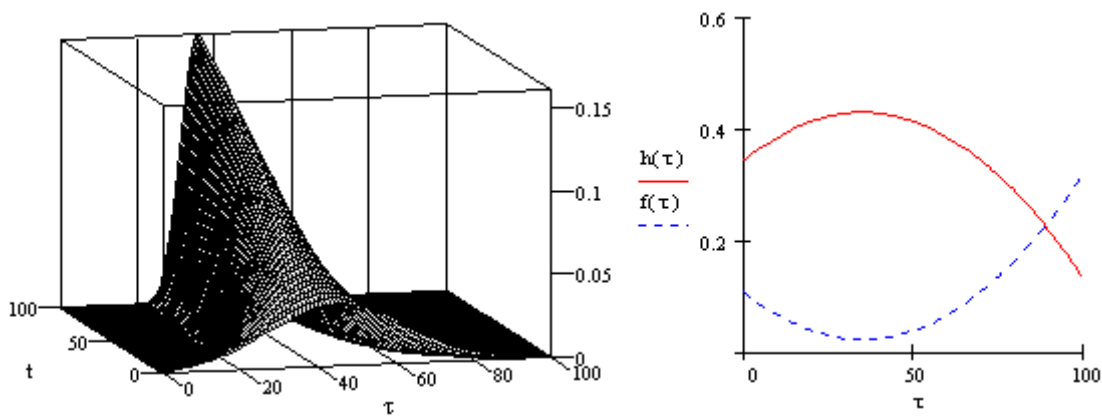

Fig. 3. Redistribution of resources due to the competition processes concentrated at 40-years age.

We think that the problem of the resource distribution control with the help of variation of $h(\tau)$ and $f(\tau)$ functions is very interesting for further investigation.

The next step will incorporate more global social and economic schemes reflecting correlation of resource distribution with respect to the age parameter and the rate of economic growth. The real parameters based on statistical data are of importance in computer simulation of real situation. Hence to fit parameters it is necessary to organize serious cooperation with economists and sociologists.